\documentstyle[psfig]{aa}
\def\zabs{$z_{\rm abs}$}
\def\zem{$z_{\rm em}$~}
\def\lya{Lyman-$\alpha$ }

\def\h2{H$_2$}
\def\hi{H{\sc i}~}

\def\nv{N~{\sc v}~ }

\def\cii{C~{\sc ii}~}
\def\ciis{C~{\sc ii*}~}
\def\mgii{Mg~{\sc ii}~}
\def\civ{C~{\sc iv}~}
\def\civa{C~{\sc iv}$\lambda$1548~ }
\def\civb{C~{\sc iv}$\lambda$1550~ }
\def\nva{N~{\sc v}$\lambda$1238~ }

\def\alii{Al~{\sc ii}~}
\def\aliii{Al~{\sc iii}~}

\def\siiis{Si~{\sc ii*}~}
\def\siii{Si~{\sc ii}~}

\def\siiv{Si~{\sc iv}~}
\def\siiva{Si~{\sc iv}$\lambda$1393~}

\def\kms{km~s$^{-1}$}
\begin{document}
%\small
%
%\thesaurus{11.17.1;11.17.4}
\title{A near-solar metallicity damped Lyman-$\alpha$ system toward the BAL
quasar Tol~1037$-$2703 
\thanks{Based on observations collected during ESO programme 65.P-0038
at the European Southern Observatory with UVES mounted on the 8.2m KUEYEN
telescope operated on Cerro Paranal, Chile}
}
\author{R. Srianand\inst{1} \& Patrick Petitjean \inst{2,3}
 }
\institute{
$^1$IUCAA, Post Bag 4, Ganeshkhind, Pune 411 007, India \\
$^2$Institut d'Astrophysique de Paris -- CNRS, 98bis Boulevard 
Arago, F-75014 Paris, France\\
$^3$UA CNRS 173 -- DAEC, Observatoire de Paris-Meudon, F-92195 Meudon
Cedex, France
}
%
%\date{\today}
\offprints{R.Srianand}
\authorrunning{Srianand, Petitjean}
\titlerunning{BAL quasar Tol 1037}
%\maketitle
\markboth{}{}

\abstract{
We report the detection of a Broad Absorption Line (BAL) outflow
in the spectrum of the \zem(Mg~{\sc ii})~=~2.201 QSO Tol 1037$-$2703 
with three main BALs at 36000, 25300 and 22300~\kms outflow velocities. 
Although the overall flow is dominated by high ionization lines 
like \nv and \civ, the gas of highest velocity shows absorption from
Mg~{\sc i}, Mg~{\sc ii} and Fe~{\sc ii}.
Covering factor arguments suggest that the absorbing complexes are
physically associated with the QSO and have transverse dimensions
smaller than that of the UV continuum emitting region
($r$~$<$~0.1~pc).  We show that the C~{\sc iv} absorption at
\zabs~=~2.082 has a covering factor $f_{\rm c}$~$\sim$~0.86 and the
absorption profile has varied over the last four years. The detection
of absorption from excited fine structure levels of C~{\sc ii} and
Si~{\sc ii} in narrow components embedded in the C~{\sc iv} trough
reveals large density inhomogeneities. IR pumping is the most likely
excitation process.
The \zabs~=~2.139 system is a moderately damped Lyman-$\alpha$ system
with log~$N$(H~{\sc i})~$\sim$~19.7. The weakness of the metal lines
together with the high quality of the data make the metallicity
measurements particularly reliable. The absolute metallicity is close
to solar with [Zn/H]~=~$-$0.26.  The $\alpha$-chain elements have
metallicities consistently solar (respectively +0.05, $-$0.02, $-$0.03
and $-$0.15 for [Mg/H], [Si/H], [P/H] and [S/H]) and iron peak
elements are depleted by a factor of about two ([Fe/Zn], [Cr/Zn],
[Mn/Zn] and [Ni/Zn] are equal to $-$0.39, $-$0.27, $-$0.49,
$-$0.30). Lines from C~{\sc i} are detected but H$_2$ is absent with a
molecular to neutral hydrogen fraction less than
8$\times$10$^{-6}$. From the ionization state of the gas, we argue
that the system is situated $\sim$few Mpc away from the QSO. High
metallicity and low nitrogen abundance, [N/Zn]~=~$-$1.40, favor the
idea that metals have been released by massive stars during a
starburst of less than 0.5~Gyr of age.  
Using the upper
limit on the C~{\sc i}$^*$ column density in two components, we obtain 
upper limits on the background temperature of 16.2 and 13.2 K respectively.
}
\maketitle

%%%%%%%%%%%%%%%%%%%%%%%%%%%%%%%%%%%

%\vskip -0.3cm
\section {Introduction}
\begin{figure*}
\centerline{\vbox{
\psfig{figure=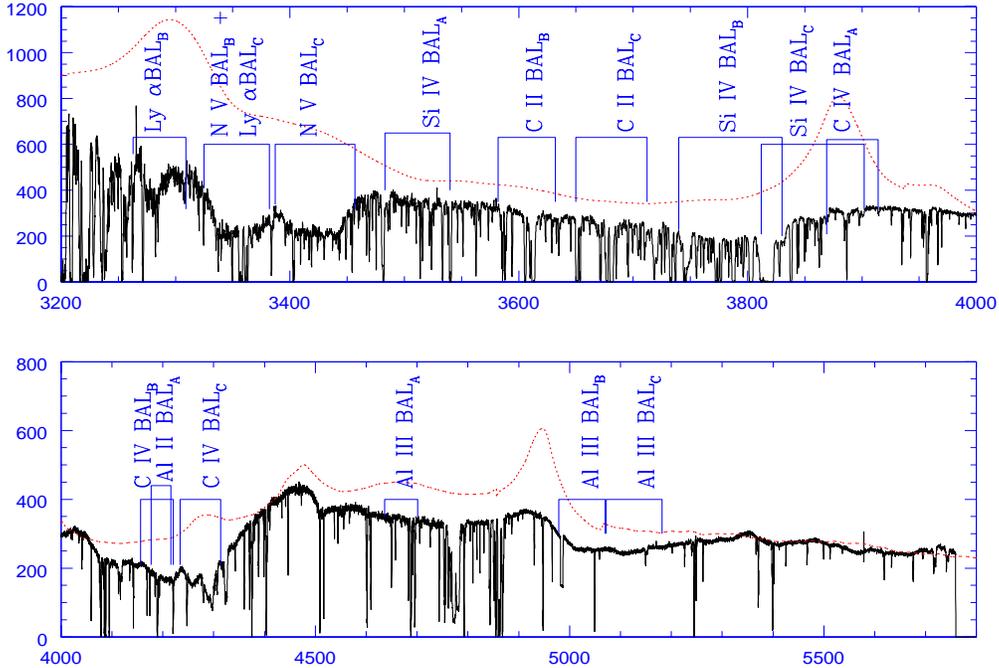,height=10.cm,width=14.5cm,angle=270}
}}
\caption[]{Portion of the Tol 1037$-$2703 spectrum. 
A high-quality UVES spectrum of Q1101$-$264 shifted in wavelength and
intensity is overplotted (dotted line). 
Position of various BALs discussed in the text are indicated. 
The subscripts 'A', 'B' and 'C' denote the \zabs~=~1.51,
1.70 and 1.75 systems respectively.
}
\label{sed}
\end{figure*}

The remarkable similarity of four \civ absorption systems at redshift
1.95~$<$~\zabs~$<$~2.14 in the spectra of Tol~1037$-$2703 (\zem~=~2.20) 
and 1038$-$2712 (\zem~=~2.33) separated by only 17.9~arcmin on the sky has 
created considerable interest in the past. However the nature and origin 
of these common absorption systems are still unclear. 
There are arguments as well as counter arguments
for (i) the absorption being produced by a super$-$cluster sitting in
front of the two quasars and (ii) part of the systems being intrinsically 
associated with the QSOs  (see Jakobsen et al. 1986, Ulrich \& Perryman 1986, 
Cristiani et al. 1987, Robertson 1987, Sargent \& Steidel 1987, Dinshaw \& 
Impey 1996, Lespine \& Petitjean 1997). 
\par\noindent
Here we use a high resolution and high S/N ratio spectrum of Tol~1037$-$2703 
of quality an order of magnitude higher than previous data to investigate the 
nature of the systems close to the quasar redshift, mainly the broad outflow
and the systems at 
$z_{\rm abs}$~$\sim$~2.082 and 2.139. As it is now established that the 
absorbing gas physically associated with the QSOs often shows some or all of:  
(i) partial coverage, (ii) excited fine-structure lines, (iii) time 
variability, (iv) broader albeit smoother profile, (v) high metal enrichment 
(see e.g. Petitjean et al. 1994, Hamann 1997, Petitjean \& Srianand 1999, 
Srianand \& Petitjean 2000), we can investigate the nature of these systems 
using these indicators.
%in more details. 
%
\par\noindent
Observations are described in Section~2. Section~3 presents the spectral 
energy distribution and accurate redshift of the QSO. We discuss the BAL flow 
in Section 4 and study the nature of the \zabs$\sim$2.082 and 2.139 \civ 
systems in 
greater detail in Section 5. The results are summarized and discussed in 
Section 6.
\section{Observations}
The Ultra-violet and Visible Echelle Spectrograph (D'Odorico et al. 2000)
mounted on the ESO Kueyen 8.2~m telescope at the Paranal observatory
has been used on April 5 to 8, 2000 to obtain high-spectral
resolution spectra of Tol~1037$-$2703. The slit width was 1~arcsec (the seeing 
$FWHM$ was most of the time better than 0.8~arcsec) and the CCDs were binned 
2$\times$2 resulting in a resolution of $\sim$45000. The total exposure time 
$\sim$9~hours was split into 1~h exposures. The data was reduced in the 
dedicated context of MIDAS, the ESO data reduction package, using the UVES
pipeline in an interactive mode. The main characteristics of the pipeline is 
to perform a precise inter-order background subtraction for science frames and
master flat-fields, and an optimal extraction of the object signal rejecting 
cosmic ray impacts and subtracting the sky at the same time.
The reduction is checked step by step. Wavelengths were corrected to 
vacuum-heliocentric values and individual 1D spectra were combined together.  
The resulting S/N ratio per pixel is of the order of 20 at $\sim$3500~\AA~ 
and 40 at $\sim$6000~\AA.
\section {BAL flow and emission redshift}
Part of the observed spectrum of Tol~1037$-$2703 is shown in
Fig.~\ref{sed}. For comparison we have also plotted the suitably 
shifted continuum of Q~1101$-264$. It is apparent that most of the 
emission lines are weak due to numerous broad absorption troughs. 
Expected positions of various absorption lines due to three 
BALs are marked. These systems are discussed in detail in 
the next Section.
\par\noindent
\begin{figure}
\centerline{\vbox{
\psfig{figure=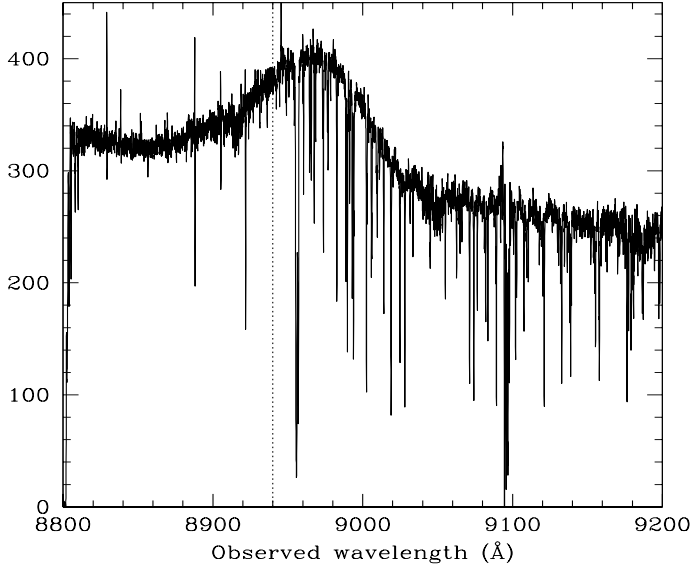,height=8.cm,width=9.5cm,angle=0}
}}
\caption[]{ The Mg~{\sc ii} emission line of Tol 1037$-$2703
at a redshift of $z$~=~2.201.
The vertical dotted line marks the expected position the emission line for the
$z_{\rm em}$~=~2.14 redshift defined by high ionization emission lines. 
}
\label{mg2}
\end{figure}
Unlike most of the other emission lines, Mg~{\sc ii} emission is
clearly seen in our red spectrum (see Fig.~\ref{mg2}).  This provides
us with the accurate measurement of the emission redshift of the QSO,
$z_{\rm em}$~=~2.201, which is important for the study of associated
systems.  Jakobsen et al. (1986) have measured $z_{\rm em}$~=~2.193
from the high excitation emission lines, which is 2400~km~s$^{-1}$
smaller than our determination. Gaskell (1982) has shown, however,
that there is a systematic difference between the redshifts derived
from broad emission-lines of different ionization levels. The mean blue-shift
of C~{\sc iv}, N~{\sc v} and possibly Lyman-$\alpha$ with respect to 
low-ionization lines as O~{\sc i}, Mg~{\sc ii} and the Balmer lines is about
600~km~s$^{-1}$ (Espey et al. 1989). The corresponding blueshift for 
Tol~1037$-$2703 is larger than this but not exceptional.
%
%
%%%%%%%%%%%%%%%%%%%%%%%%%%%%%%%%%%%%%%%%
\section {Broad absorption systems}
%%%%%%%%%%%%%%%%%%%%%%%%%%%%%%%%%%%%%%%%
%
\subsection{\zabs~=~1.5181}
\begin{figure}
\centerline{\vbox{
\psfig{figure=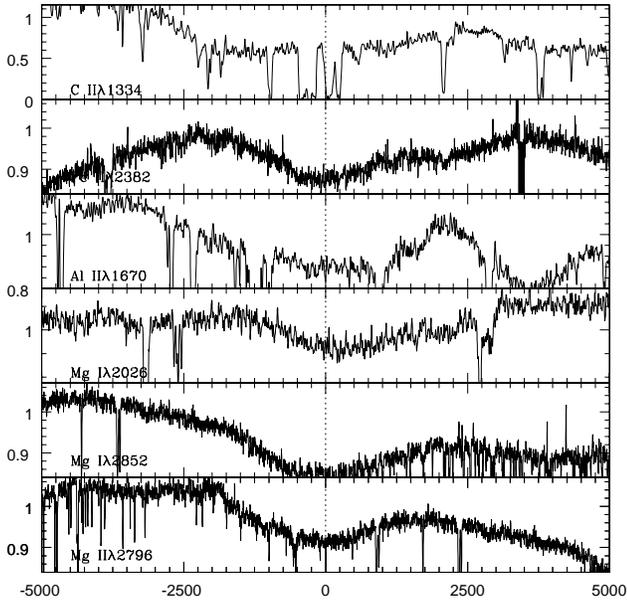,height=9.cm,width=9.cm,angle=0}
}}
\caption[]{Velocity plot of broad absorption lines from
the \zabs~=~1.5181 system.}
\label{bal1}
\end{figure}
This system is identified by broad absorption lines of 
Mg~{\sc ii}$\lambda$2800 and Mg~{\sc i}$\lambda2852$. 
Presence of Fe~{\sc ii}$\lambda2382$, Fe~{\sc ii}$\lambda$2600 and 
possibly Mg~{\sc i}$\lambda2026$ in the red spectrum (see Fig.~\ref{bal1}) 
confirms the identification. Note  that the unidentified line in the low 
resolution spectrum of Ulrich \& Perryman (1986) correspond to Fe~{\sc ii} 
at this redshift (see their Fig.~1). The \mgii absorption trough has 
$FWHM \sim2400$ \kms~ and ejection velocity 36000 \kms~ with respect to the 
QSO. Absorption lines due to Al~{\sc ii}$\lambda1670$ and 
Si~{\sc ii}$\lambda1304$ could be present. The Al~{\sc ii} line is probably 
blended with \civ absorption at \zabs~=~1.7 (see below). 
Si~{\sc ii}$\lambda$1526 and \civ absorptions are redshifted on top of the 
QSO \lya+N~{\sc v} emission line and could possibly be responsible for the 
weakness of the emission feature. The expected position of 
C~{\sc ii}$\lambda$1334 coincides with the broad feature at 3360 \AA, together
with \lya and \nv BALs at \zabs~=~1.76 (see below). Blending with the 
intervening \lya absorption lines and other broad absorption troughs prevents 
systematic estimate of column densities. 
\par\noindent
The so-called low ionization Mg~{\sc ii} BALs are seen in 15\% of the
BALQSOs detected in optical surveys (Voit et al. 1993). All the
known Mg~{\sc ii} BAL systems have strong absorption due to \civ, \alii and
\aliii; the absorption due to \aliii being prominent compared to that of
\alii. The profiles of low ionization BALs are always shallower and
narrower than that of high ionization transitions (Voit et al.
1993). Becker et al. (2000) have suggested that the fraction of
BALs amongst radio selected QSOs are high and $\sim50\%$ of them show
absorption due to low ionization species. It is interesting to
note that no BAL due to Mg~{\sc i} has been mentioned up to now
(see however de~Kool et al. 2001).  
\par\noindent
After carefully taking into account the atmospheric features, the 
Mg~{\sc i}$\lambda$2852 
rest equivalent width is $w_{\rm r}$~=~0.78~\AA~ and $FWHM$~=~ 1523 \kms.  
Correspondingly, $w_{\rm r}$(Mg~{\sc ii}) is 2.35~\AA. This
gives $N$(\mgii)$\ge4\times10^{13}$ and $N$(Mg~{\sc i})$\ge1\times10^{14}$~
cm$^{-2}$. 
As the ionization potential of Mg~{\sc i} is less than 13.6 eV,
even if hydrogen is completely neutral, $N$(Mg~{\sc ii}) is expected
and observed to be larger than $N$(Mg~{\sc i}). This strongly suggests that
the absorbing gas does not completely cover the background source. From the 
residual intensities we estimate that the covering factor could be 
as low as $\sim0.1$ which would be consistent with the strength of the 
Fe~{\sc ii} lines.
The Mg~{\sc ii} absorption is redshifted in a region devoid of any
emission line.  This means the absorbing gas has a projected size
of less than 0.01 pc (10\% of the typical size of the UV continuum emitting
region). It must be noted that the presence of such low covering
factor in BAL flows has already been mentioned (Hamann
et al. 1997, Telfer et al. 1998, Srianand \& Petitjean 2000).
%
%%%%%%%%%%%%%
\subsection{\zabs~=~1.7018 and 1.7569 }
%%%%%%%%%%%%%%
\begin{figure}
\centerline{\vbox{
\psfig{figure=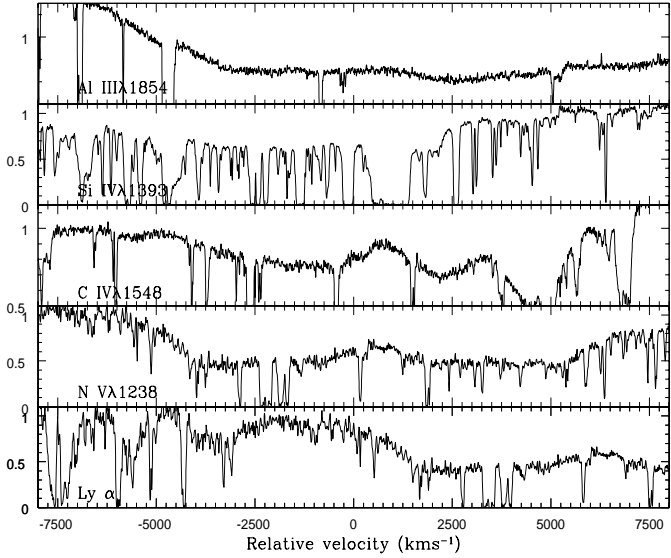,height=8.cm,width=10.cm,angle=270}
}}
\caption[]{Velocity plot of broad absorption lines in the twin
BAL system at \zabs~=~1.7018 and 1.7569.
}
\label{bal2}
\end{figure}
The twin system is detected mainly through a strong \nv through seen at
$\lambda_{\rm obs}~\sim~3400~$\AA. From Fig.~\ref{bal2} it is apparent
that \civ absorption is present in both components. Although not excluded, it 
is unclear whether \siiv and Al~{\sc iii} absorptions are present.
It is interesting to note that the bottom of the \nv absorption is flat for 
both components but without going to zero. 
Continuum in this wavelength range is not well defined however
(see Fig.~\ref{sed}).Therefore, by fitting the continuum locally we
find that the residual intensity is $\sim0.5$, suggesting the covering factor 
of \nv could be close to 0.5. If we normalise the spectrum by fitting
the regions free from absorption to the Q~1101$-264$ spectrum, we derive a 
covering factor of $\sim 0.7$. As the \nv lines are saturated 
we obtain tentative lower limits on the N~{\sc v} column densities, 
$\ge3.4\times10^{16}$ and $\ge4.0\times10^{16}$~cm$^{-2}$ respectively for 
the two components. The $FWHM$ are $\sim 4000$ \kms and $\sim6000$ \kms~ 
respectively. These components are at an ejection velocity of 25300 and 
22320 \kms with respect to the QSO redshift defined by the \mgii
emission. 
\par\noindent
The absence of absorption due to singly ionized species, the weakness of the
H~{\sc i} Lyman-$\alpha$ line and the fact that N~{\sc v} is stronger than 
C~{\sc iv} suggest that ionization conditions in these systems are similar 
to those of standard high-ionization BALs. Assuming the nitrogen abundance 
to be ten times solar and \nv to be the dominant nitrogen species we estimate 
a lower limit on the total hydrogen column density of 3.9$\times10^{19}$ 
and 4.2$\times10^{19}~{\rm cm}^{-2}$ for the two components.
The observed upper limit on $N$(H~{\sc i}) at \zabs~=~1.7018
is 10$^{15} ~{\rm cm}^{-2}$ implying log~$U\ge-1.0$ (Hamann 1997). 
\subsection{\zabs~=~2.082 }
\begin{figure}
\centerline{\vbox{
\psfig{figure=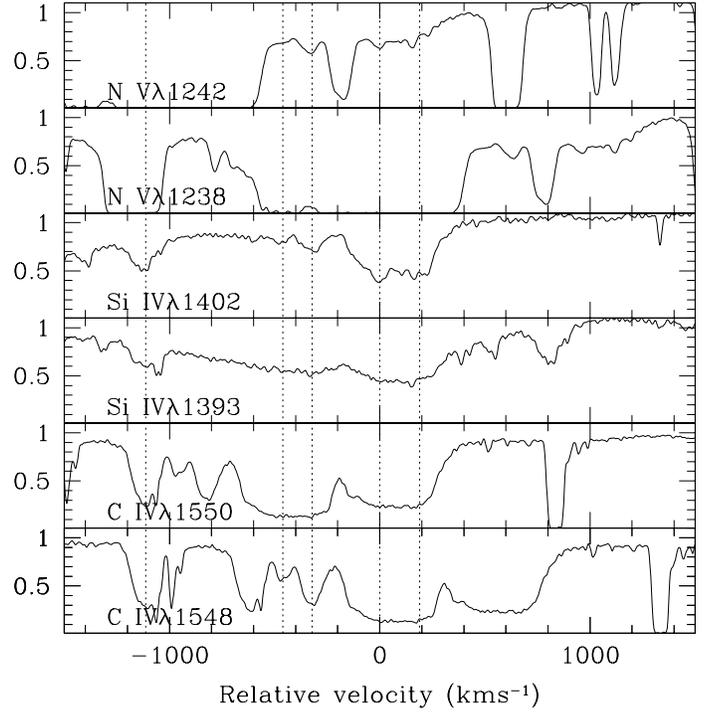,height=10.cm,width=10.cm,angle=0}
}}
\caption[]{Velocity plot of high ionization lines from
the \zabs~=~2.082 system. The vertical dotted lines mark some of the
components discussed in the text. 
}
\label{hmbal}
\end{figure}
\begin{figure}
\centerline{\vbox{
\psfig{figure=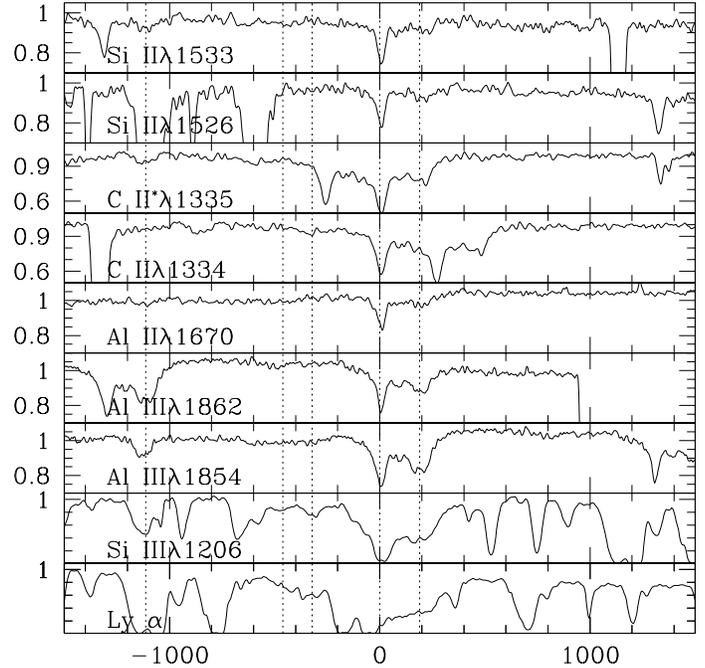,height=10.cm,width=10.cm,angle=0}
}}
\caption[]{Velocity plot of low ionization lines from
the \zabs~=~2.082 system. The vertical dotted lines mark some of the
components discussed in the text. The main narrow component is at 
$z_{\rm abs}$~=~2.0819.
}
\label{lmbal}
\end{figure}
This system has induced considerable interest among astronomers as
there is a similar, albeit diffuse, broad absorption trough 
at the same redshift in the spectrum of QSO Tol 1038$-$2712
which is at a projected separation of 
17.9~arcmin on the sky (or $\sim$4.5$h^{-1}$~Mpc at $z$~=~2) from 
Tol~1037$-$2703. Whether the gas is
associated with the QSO or with an intervening absorption complex is
unclear however (see Jakobsen et al. 1986, Ulrich \& Perryman 1986, Cristiani
et al. 1987,  Robertson 1987, Sargent \& Steidel 1987, Dinshaw \&
Impey 1996, Lespine \& Petitjean 1997). 
Absorption profiles of high-ionization lines 
are shown in Fig.~\ref{hmbal}. \siiva is blended with \civ of the BAL system at
\zabs~=~1.7569 (see Fig.~\ref{bal2}) and \nva is blended with Lyman-$\alpha$ 
at \zabs~=~2.139 (see Fig.~\ref{met}).  
The \civ absorption at $v~=~0$~\kms~ is broad and unresolved even at our
resolution. The relative strength of \civa and \civb absorption lines suggests 
that the lines are moderately saturated and the gas covers only part of the
background source. As these lines are redshifted in a region devoid of 
emission lines, the projected size of the absorbing gas has to be less than 
the typical dimension of the region emitting the continuum. 
\par\noindent
This component also shows absorption due to low ionization lines such as 
Al~{\sc ii}, Si~{\sc ii}, C~{\sc ii}, Si~{\sc iii}, Al~{\sc iii} and 
absorption due to fine-structure lines of Si~{\sc ii} and C~{\sc ii} (see 
Fig.~\ref{lmbal}). The expected position of Mg~{\sc ii}$\lambda$2797 
unfortunately coincides with a small gap in our UVES spectrum. The detected 
low ionization lines have a complex structure with a strong narrow component 
at $z_{\rm abs}$~=~2.0819 superimposed to a diffuse broader profile.
\subsubsection {Partial coverage}
\begin{figure}
\centerline{\vbox{
\psfig{figure=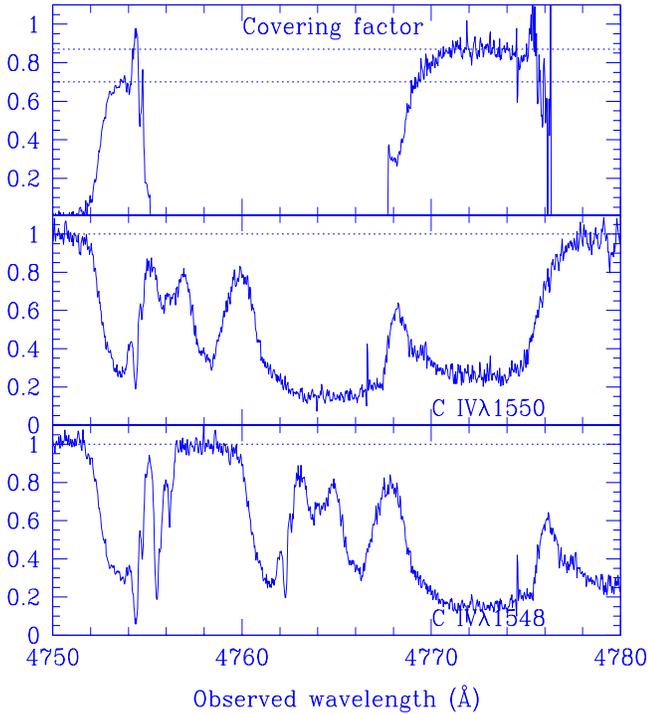,height=10.cm,width=9.cm,angle=0}
}}
\caption[]{Covering factor estimated from the \civ absorption at \zabs~=~2.082
(top panel). The lower and middle panels give the \civa and \civb profiles 
(suitably shifted). 
}
\label{c4cf}
\end{figure}
We estimate the covering factor, $f_{\rm c}$, of the \civ gas using the 
method described in Srianand \& Shankaranarayanan (1999).
In the case of unblended doublets one can write
\begin{equation}
f_{\rm c}~=~{1+R_2^2-2R_2 \over 1+R_1-2R_2},
\end{equation}
where $R_1$ and $R_2$ are the residual intensities in the
first and second lines of the doublet. 
Typical error in the estimated covering factor is equal to
rms in the normalised continuum. As the \civ absorption lines from
this system are not contaminated by any broad absorption trough (see
Fig.~\ref{sed}) our continuum estimation, based on spline fit,
in this wavelength range is reliable. The lower and middle panels of 
Fig.~\ref{c4cf} gives the normalised \civa and \civb
profiles on a velocity scale. The top panel gives the estimated covering 
factor. The \civ absorption at $v\sim0$~\kms~ (i.e.  $\lambda\sim$ 4773~\AA) 
is consistent with a covering factor $f_{\rm c}$~=~0.86$\pm$0.03. It is to 
be noted that \civa of this component has some contamination from \civb 
of the two narrower \civa components at $\lambda\sim 4764$ and 4766~\AA. 
Taking into account this contamination would decrease the covering factor.
Moreover, the fact that the \civb components are not seen superimposed onto 
the flat bottom of the \civa profile (at $\lambda\sim4773$~\AA) strongly 
supports the idea that the \civa line is optically thick
at this position and $f_{\rm c}$~=~0.86. {\sl This implies as 
well that the three clouds (with central} \civa {\sl wavelengths at 
$\lambda\sim$ 4764, 4766 and 4773~\AA) must cover the same part of the 
background source}. This could be a consequence of the absorption arising 
through a strongly collimated flow.
\par\noindent
The estimated covering factor for \civ is constant within the absorption line
profile. This suggests that there is no large variations of the 
line of sight velocity field across the absorbing cloud. The estimated
covering factor for the component at $v\sim-1110$ \kms (i.e.
$\lambda\sim$4754~\AA) is $f_{\rm c}\sim 0.7$. The distinct narrow component 
in the red wing showing complete coverage probably corresponds to a normal
intervening system.
\begin{figure}
\centerline{\vbox{
\psfig{figure=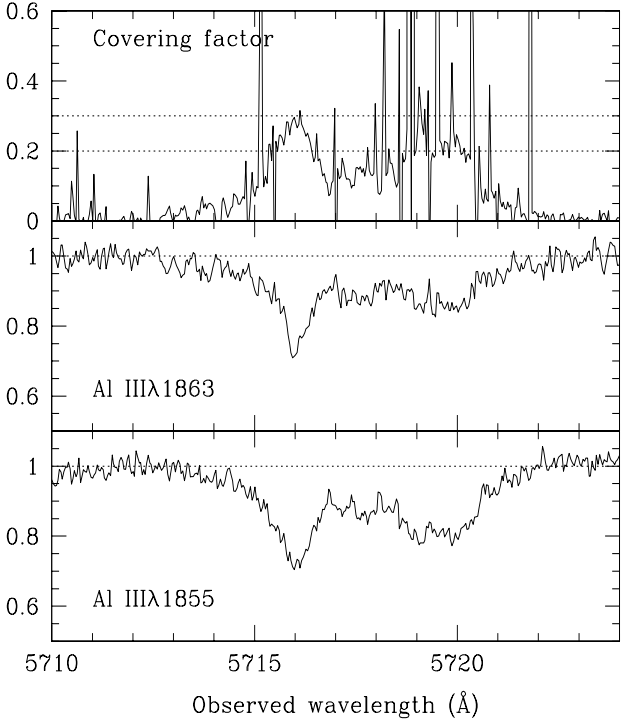,height=10.cm,width=9.cm,angle=0}
}}
\caption[]{Covering factor estimated from the \aliii absorption at 
\zabs~=~2.082 (top panel). Lower and middle panels give respectively the 
profiles of Al~{\sc iii}$\lambda1855$ and Al~{\sc iii}$\lambda1863$ (suitably
shifted). 
}
\label{al3cf}
\end{figure}
\begin{figure}
\centerline{\vbox{
\psfig{figure=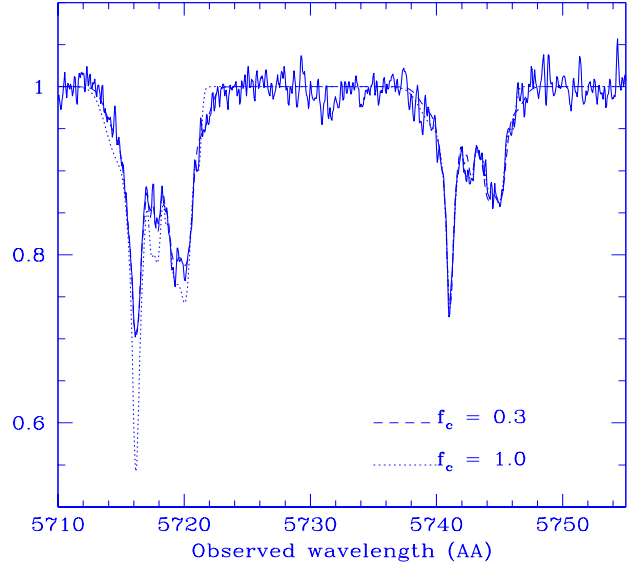,height=8.cm,width=9.cm,angle=0}
}}
\caption[]{Voigt profile fits to the \aliii doublets. The observed
(solid line) and model profiles for complete coverage (dotted line)
and with 30\% coverage (dashed line) are presented.}
\label{al3fit}
\end{figure}
\par\noindent
The same analysis is done for the \aliii doublet (see Fig.~\ref{al3cf})
which is not contaminated by any broad absorption trough.
In this wavelength range too, the continuum is well approximated by
spline fitting.
As discussed before the profile shows a narrow component on top of a broad 
absorption structure. Within the uncertainties, the covering factor 
is in the range $\sim0.2-0.3$ (see Fig.~\ref{al3fit}). It is well known 
however that the blending of weak unresolved lines can mimic partial coverage 
(Lespine \& Petitjean 1997). In order to ascertain the above result, we
have fitted the Al~{\sc iii} doublet with multiple Voigt profile components
taking into account the partial coverage. 
The best fit result is presented in Fig.~\ref{al3fit}. It is apparent from the
figure that, at least for the strongest component, partial covering factor 
($f_{\rm c}$~$\sim$~0.3) is needed. This confirms that the gas responsible for
the \zabs~=~2.082 system covers the background source only partially. The 
estimated covering factor varies from one species to the other, being larger 
for higher excitation lines (see also Petitjean \& Srianand 1999; Srianand \&
Petitjean 2000). 
\subsubsection{Excited fine-structure lines}
It can be seen on Fig.~\ref{lmbal} that absorption lines from low-ionization
species (C~{\sc ii}, Si~{\sc ii} and Al~{\sc ii}) are present in a well
defined narrow component at $z_{\rm abs}$~=~2.0819. The presence of
strong associated \siiis and \ciis lines suggests either that the electron
density of the gas is large or that the gas is very close to an IR emitting
source. 
\par\noindent
The observed equivalent widths and profiles of the two lines  
C~{\sc ii}$\lambda1334$ and C~{\sc ii}$^*$$\lambda1335$ are similar. Since 
their rest wavelength and oscillator strengths (0.1278 and 0.1149 
respectively) are nearly identical, \cii and \ciis have about the same column 
densities, if the lines are not saturated.
% and partial covering factor. 
If we assume electron collisions are responsible for the excitation, the 
electron density $n_{\rm e}$ is then given by,
\begin{equation}
n_e\sim75~T_4^{0.5}/[{\rm exp}({-91.25/ T})-1]~{\rm cm^{-3}},  
\end{equation}
where $T_4$ is the kinetic temperature in units of $10^4$~K. The relevant
cross-sections are taken from Bahcall \& Wolfe (1968). 
\par\noindent
From the strength of Si~{\sc ii}$\lambda1526$ and Si~{\sc ii$^*$}$\lambda1533$,
at $z_{\rm abs}$ = 2.0819, we derive that the ratio
$\beta$~=~$N$(Si~{\sc ii})/$N$(Si~{\sc ii$^*$}) is in the
range 1.58~$<$~$\beta$~$<$~2.36 for a covering factor in the range 0.3$-$1.0.  
Note that the above range also includes Voigt profile fitting
errors in estimating the individual column densities. Assuming 
excitation and de-excitation are due to electron collisions we obtain
\begin{equation}
n_{\rm e}~=~{1.28\times10^3~T_4^{0.5}\over {[2\beta {\rm exp}(-413/T)-1]}}
{\rm cm^{-3}}.
\label{eqn2}
\end{equation}
We therefore derive that the electron
density is in the range 360$-630~{\rm cm}^{-3}$ for an assumed temperature
of 10$^4$K. This is inconsistent with the density derived from \cii.
This implies that another excitation mechanism is at play. Note that although
it is difficult to estimate the $N$(Si~{\sc ii})/$N$(Si~{\sc ii$^*$}) column 
density ratio in the broader component, it is consistent with that of the 
narrow component.
\par\noindent
Indeed excited levels can be populated by IR radiation emitted by
the QSOs. In that, case,
\begin{equation}
{N(X^*)\over N(X)}~=~2\times\bigg({\bar n_\lambda \over 1+\bar n_\lambda}\bigg)
\end{equation}
Where,$ X$, is either \cii or \siii and $\bar n_\lambda$ is given by
\begin{equation}
\bar n_\lambda~=~{I(\nu)\lambda^3\over 8\pi hc}
\end{equation}
where, $I(\nu)$, in erg cm$^{-2}$ s$^{-1}$ Hz$^{-1}$, is the flux 
energy density integrated in all directions.
From the observed column density ratios we get,
\begin{equation}
\bar n_\lambda({\rm \ciis})~=~(2-4) \bar n_\lambda({\rm Si~{\sc ii}^*}).
\end{equation}
Such a flux energy density ratio can be achieved by an IR spectrum, 
${I(\nu)\propto \nu^\alpha}$, with $\alpha$ in the range 2 to 2.5. Even 
though the IR spectrum of Tol~1037$-$2703 is not known, the required 
spectral index is typical of what is seen in high redshift QSOs
(e.g. Elvis et al. 1994). 
\par\noindent
Therefore the dominant excitation process is probably pumping by the 
infrared ambient radiation field and it is not possible to estimate
the electronic density from Eqs.~1 and 2. Srianand \& Petitjean (2000) 
reached a similar conclusion to explain the excitation of fine$-$structure 
lines 
in the \zabs~=~3.8931 system toward APM~08279$+$5255. 

\subsubsection {Variability}
\begin{figure}
\centerline{\vbox{
\psfig{figure=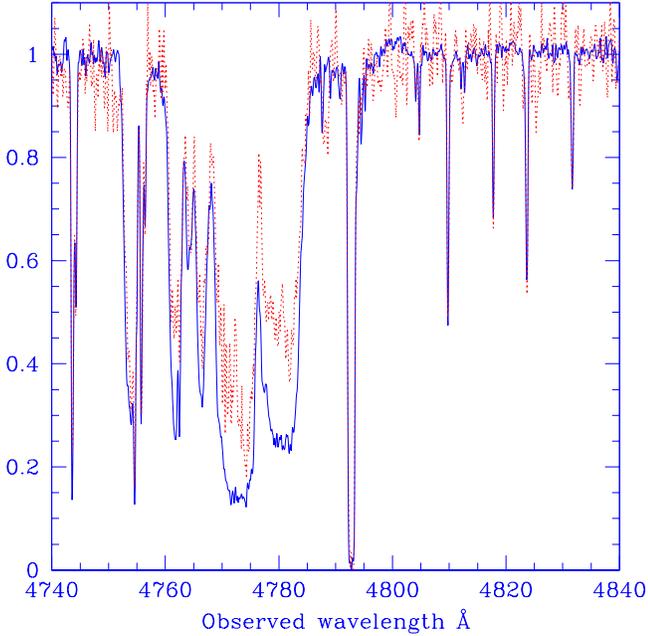,height=9.cm,width=9.cm,angle=0}
}}
\caption[]{Time variability of \civ absorption. The continuous line
corresponds to our UVES data and the dotted line is the CASPEC spectrum from
Lespine \& Petitjean (1997). The resolution of the UVES spectrum is degraded
to match that of the CASPEC data. 
}
\label{varc4}
\end{figure}
\begin{figure}
\centerline{\vbox{
\psfig{figure=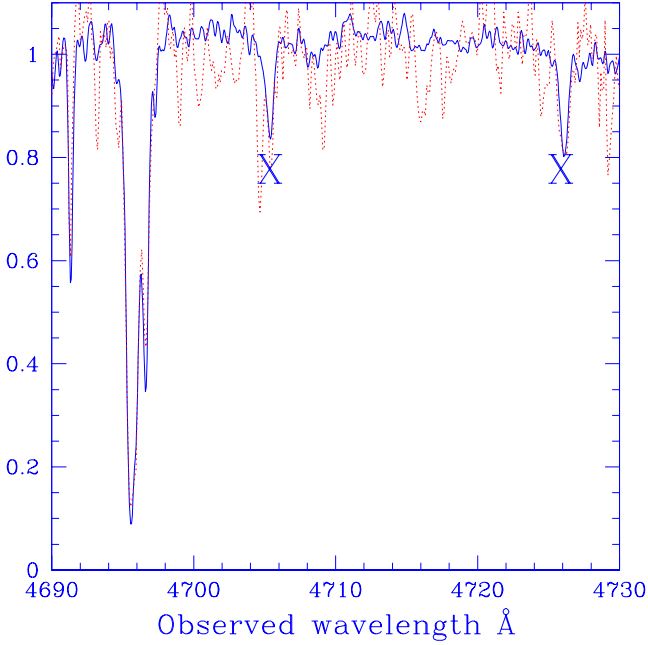,height=9.cm,width=9.cm,angle=0}
}}
\caption[]{Time variability of Si~{\sc ii} absorption lines (marked with 'X'). 
The solid line corresponds to the UVES data and the dotted line is the
CASPEC data from Lespine \& Petitjean (1997).}
\label{si2var}
\end{figure}

If the presence of excited fine-structure lines and partial covering factor
is due to the density of the gas being high, then it is possible, in
principle, to constrain the density from variability observations. 
To investigate this, we use the high-resolution spectrum obtained 
with the instrument CASPEC on the ESO 3.6~m telescope. This spectrum is 
the combination of data obtained during two observing runs in December 1992 
\& April 1994 (see Lespine \& Petitjean 1997). This corresponds to an 
elapsed time of $\sim2$ yrs in the rest frame of the QSO compared to the
new UVES data. The CASPEC data has slightly lower resolution and has inherent 
problems with background subtraction due to small inter-order spacing 
in the blue. We therefore have convolved the UVES spectrum with a Gaussian 
profile to match the width of the narrow lines in the CASPEC spectrum. We have 
corrected for the background inaccuracy by matching the depth of the
narrow lines due to intervening systems which are present close to the
\civ absorption at $\lambda$4744 and $\lambda$4893 (see Fig.~\ref{varc4}). 
From Fig.~\ref{varc4}, it is clear that, while the narrow lines in both 
spectra are identical, the broad lines covering the observed wavelength range 
4750$-$4790~\AA~ have become stronger with time. In addition, the shape of 
the broadest component has changed. This variability is confirmed in
a MMT spectrum taken in 1994 (Dinshaw \& Impey 1996). We do not see any 
variation in the relative position of the different components suggesting that
acceleration is small. Note however that the narrow Si~{\sc ii} absorption 
lines did not vary (see Fig.~\ref{si2var}). The equivalent width 
($w_{\rm r}$~=~0.18~\AA) measured by Dinshaw \& Impey (1997) is
also consistent with what we measure. 
\par\noindent
If the variations are due to changes in the optical depth, then the 
variability time scale gives the recombination time scale. If the gas is 
highly ionized so that $n_{\rm i+1}$~$>$~$n_{\rm i}$, then the recombination 
time scale is given by,
\begin{equation}
t_{\rm r}~=~\bigg( {n_{\rm i}\over
n_{\rm i+1}}\bigg)~(n_{\rm e}\alpha_{\rm i+1})^{-1}\label{eqn3}
\end{equation}
where, $n_{\rm i}$ and  $n_{\rm i+1}$ are densities of species ionized $i-1$ 
and $i$ times
and $\alpha_{\rm i+1}$ is the recombination rate to the $i$ state (see e.g. 
Hamann et al. 1997). If the gas is less ionized, 
$n_{\rm i}$~$>$~$n_{\rm i+1}$, then,
\begin{equation}
t_{\rm r}~=~(n_{\rm e}\alpha_{\rm i})^{-1}.\label{eqn4}
\end{equation}
\par\noindent
Assuming \civ is the dominant ionization species and using the
recombination cross-section for \civ given by Aldrovandi \& P\'equignot (1986)
and P\'equignot et al. (1991), the electron density, $n_{\rm e}$, is given by 
\begin{equation}
n_{\rm e}~\ge~4.2\times10^3~T_4^{0.8}~t_{\rm yr}^{-1}~~{\rm cm}^{-3}. 
\label{eqn5}
\end{equation} 
Here, $T_4$ and $t_{\rm yr}$ are, respectively, the kinetic temperature in 
units of 10$^4$~K and the elapsed time in years. It is therefore apparent that
whatever the temperature of this gas is, the electronic density is quite large.
\par\noindent
\begin{figure}
\centerline{\vbox{
\psfig{figure=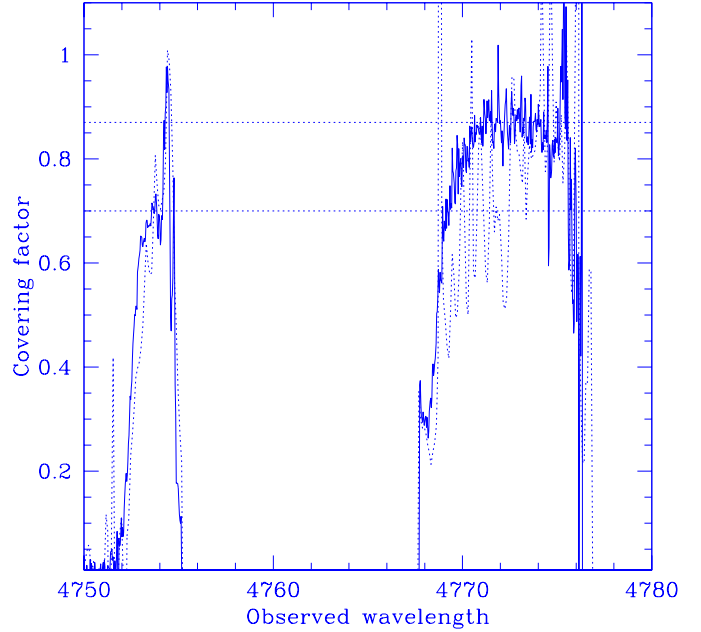,height=9.cm,width=9.cm,angle=0}
}}
\caption[]{Time variability of the covering factor of the \civ  absorption 
lines. The solid line corresponds to the UVES data and the dotted line is the
CASPEC spectrum from Lespine \& Petitjean (1997).}
\label{cfvar}
\end{figure}
\par\noindent
The comparison of covering factors estimated at two epochs (see
Fig.~\ref{cfvar}) suggests that there could be some variation from one
epoch to the other. However the S/N ratio in our CASPEC spectrum,
$\sim 10$, implies an error of up to 0.1 in the covering factor
determination. Within this uncertainty we cannot ascertain the variation
in the covering factor between the two epochs.  Also there is no
statistically significant difference in the covering factor of \civ
absorption between UVES data and MMT data of Dinshaw \& Impey (1996).
\par\noindent
%
%(see Fig.~\ref{cfvar}). 
%
%Therefore, 
It is possible that the relative increase in the \civ column density 
between years 1994 and 2000, $\ge1.2$, corresponds to a change 
of the C~{\sc iv}/C ratio and therefore a change in the ionization parameter. 
From the models by Hamann (1997) we derive that the gas has an ionization 
parameter log~$U$ of the order of $-1$ and that the increase in 
$N$(C~{\sc iv}) corresponds to a decrease of the ionization parameter by a 
factor larger than 2.
%
%\par\noindent
% 
This cannot be due to changes in the ionizing flux from the quasar as a change
by $\sim$1~mag would have been noticed. It is also difficult to imagine that 
the distance of the cloud from the central ionizing source has increased by a 
factor 1.4. Thus the observed variation in the \civ profile could be
either due to some non-equilibrium process or due to changes in the 
internal dynamics of the cloud. In order to understand the origin
of the variability it is important to have a high resolution spectroscopic 
monitoring of this system.
\section{The \zabs~=~2.139 system}
\begin{figure}
\centerline{\vbox{
\psfig{figure=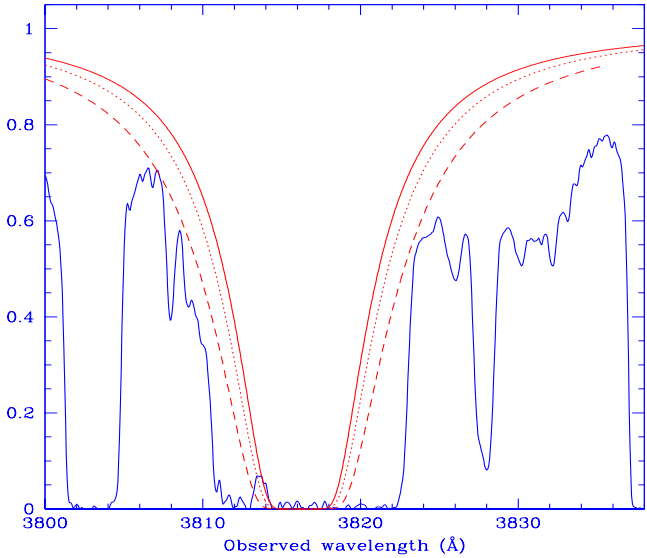,height=8.cm,width=9.5cm,angle=0}
}}
\caption[]{Profile of the \lya absorption line in the
\zabs~=~2.139 system. As this part of the spectrum is affected
by the BAL absorption the continuum was normalised using 
the spectrum of Q~1101-264. The solid, dotted and 
dashed curves are Voigt profiles for log $N(H~{\sc i})$ = 19.6, 
19.7 and 19.85 respectively. The residual flux at $\lambda$~=~3814~\AA~
suggests that log $N(H~{\sc i})\simeq$19.7.
}
\label{lya}
\end{figure}
\begin{figure}
\centerline{\vbox{
\psfig{figure=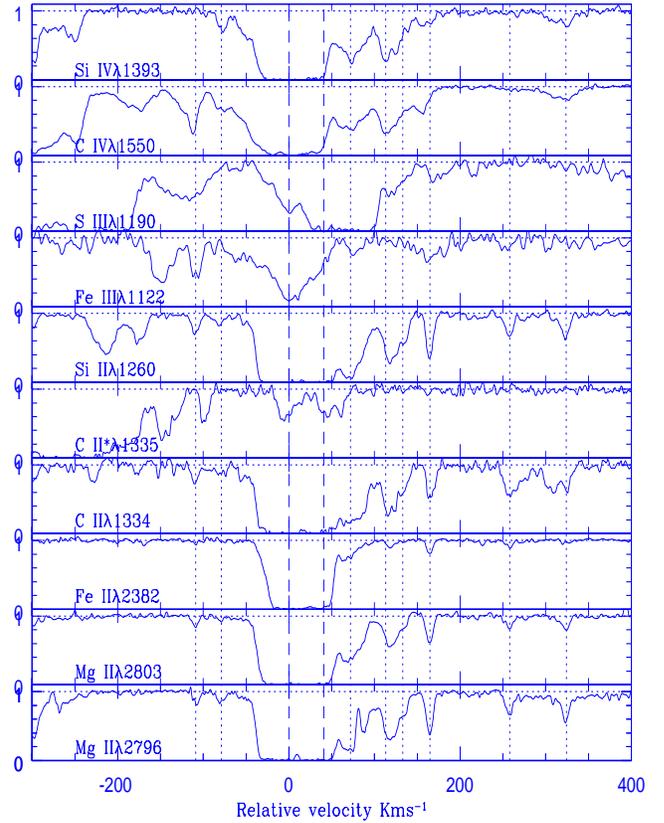,height=12.cm,width=9.5cm,angle=0}
}}
\caption[]{Absorption profiles of some of the standard
absorption lines in the \zabs~=~2.139 system. The dotted
lines show the position of the weak narrow components.
The dashed lines show the position of low ionization
components discussed in the text.}
\label{met}
\end{figure}
\subsection{Description of the system}
Lespine \& Petitjean (1997) derived an upper limit on the \hi column density, 
log~$N$(H~{\sc i})~$<$~4$\times 10^{19}~{\rm cm^{-2}}$ from the absence of 
damping wings. As can be seen from Fig.~\ref{sed}, the \lya absorption from 
this system is blended with the \nv broad absorption at \zabs~=~2.082 system as
well as the \siiv absorption from the \zabs~=~1.72 twin BAL systems.  This
means that the continuum is badly constrained. We have normalized the
Tol~1037$-$2703 spectrum using the UVES data on Q~1101$-$264 as shown
in Fig.~\ref{sed}. Although the possible damping wings are lost in the
BALs, the residual flux at $\lambda$~$\sim$~3814~\AA~ (see Fig.~\ref{lya}) 
constraints the column density well. This feature is real as it
is clearly visible in the two individual UVES exposures  we have.
We find that 19.6~$<$~log~$N$(H~{\sc i})
\par
\begin{figure}
\centerline{\vbox{
\psfig{figure=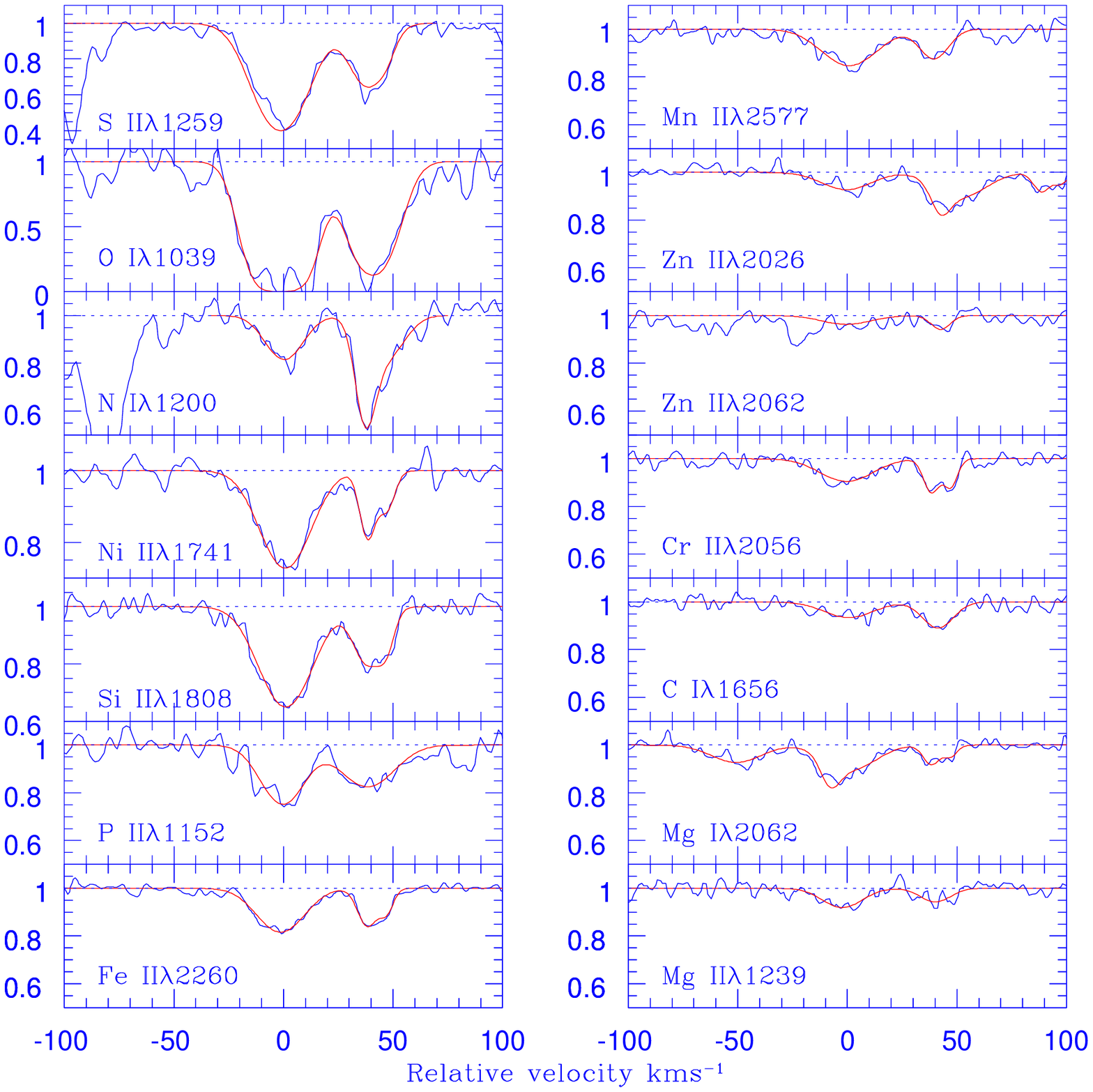,height=12.cm,width=10.cm,angle=0}
}}
\caption[]{Absorption profiles of few transitions in the \zabs~=~2.139
system. The Voigt profile fits are over-plotted.}
\label{met1}
\end{figure}

\noindent
The Mg~{\sc ii}$\lambda$2797 absorption profile shows a strong
component with narrow and weak satellites (marked with vertical dotted
lines in Fig.~\ref{met}) seen on both sides and spreading in total
$\sim$430 \kms.  This is a typical characteristic of the intervening
Mg~{\sc ii} systems. Interestingly, some of the narrow components 
exhibit Si~{\sc ii}, C~{\sc ii} and Fe~{\sc ii} absorption lines 
with weak associated absorption lines of Si~{\sc iv} and C~{\sc iv}.
When weaker transitions are considered, the strong feature splits into two 
main components of similar characteristics at $z_{\rm abs}$~=~2.1390 and 
2.1394 (marked with dashed vertical lines in Fig.~\ref{met}).
There are absorption lines at the expected positions of Fe~{\sc
iii}$\lambda1122$ and Si~{\sc iii}$\lambda1190$. As these lines 
could be blended with intervening \lya absorption lines, 
we 
derive upper limits on the corresponding column densities.
The two strong components show detectable absorption from
C~{\sc i}, N~{\sc i}, P~{\sc ii}, Mg~{\sc i}, Mg~{\sc ii}, S~{\sc ii},
Ni~{\sc ii}, Zn~{\sc ii}, Cr~{\sc ii}
(see Fig.~\ref{met1}). 
As the above lines are weak, and in most of the cases multiplets are present, 
we could obtain reliable estimate of the column densities (see Table
\ref{tab1}). 
The fact that all the saturated lines go to zero and that Voigt profiles 
could be fitted consistently to the multiplets shows that the absorbing cloud 
completely covers the background source. 
This, together with the absence of absorption line variability
and the presence of absorption due to C~{\sc i} strongly suggests
that the physical conditions in this gas are very much similar to those
prevailing in typical intervening damped Lyman-$\alpha$ system (see below).
\par\noindent
Based on the absence of absorption from molecular hydrogen in the Lyman-band,
we derive an upper limit of 10$^{14}$~cm$^{-2}$ for the \h2 column 
density in both components and therefore, the molecular fraction 
$f$~=~2$N$(H$_2$)/(2$N$(H$_2$)~+~$N$(H~{\sc i}))~$<$~8~10$^{-6}$.
As shown by Petitjean et al. (2000), such low value is probably consequence
of high temperature ($T$~$>$~3000~K) in diffuse gas.
\par\noindent
We estimate the upper limit on $N$(C~{\sc i$^{*}$}) using a combined 
spectrum obtained by stacking the regions centered at rest 
wavelengths 1277.28, 1277.51, 1329.08, 1329.10, 1329.12, 1560.68, 
1560.71, 1656.26, 1657.38 and 1657.91~\AA. There are coincident 
absorption features at the expected position of C~{\sc ii$^*$} for 
both components at \zabs~=~2.1390 and 2.1394 (see Fig~\ref{met}). 
The C~{\sc ii$^*$} absorption at \zabs~=~2.1390 is probably (at 
least partly) blended with C~{\sc ii}$\lambda$1334 from a 
narrow component at 258~\kms~ seen in Mg~{\sc ii}. The C~{\sc ii$^*$} 
absorption at \zabs~=~2.1394 has two sub-components. The red one
is blended with C~{\sc ii}$\lambda$1334 at 323~\kms, the other one at 
\zabs~=~2.1394 is free of blending. From the fact that the 
C~{\sc ii}$\lambda1036$ line is saturated we obtain a lower limit 
$N$(C~{\sc ii})~$>$~3$\times10^{14}$~cm$^{-2}$. 
\begin{table}
\caption{Column densities (in cm$^{-2}$) and metallicities relative to solar
}
\begin{tabular}{lccc}
\hline
\multicolumn{1}{c}{Species}&\multicolumn{1}{c}{$z_{\rm abs}$~=~2.1390} & 
\multicolumn{1}{c}{\zabs~=~2.1394}&\multicolumn{1}{c}{[X/H]$^{\rm a}$}\\
\hline
H~{\sc i}    & \multicolumn{2}{c}{$5\times10^{19}$} & \\
H$_2$        & \multicolumn{2}{c}{$<$10$^{14}$} & \\
C~{\sc i}    & 3.26$\pm0.17\times10^{12}$ & 3.05$\pm0.17\times10^{12}$ &  \\
C~{\sc i$^*$}& $\le 3\times10^{12}$       & $\le 2\times10^{12}$ & \\
C~{\sc ii$^*$}&         .....             & 2.03$\pm0.01\times10^{13}$ & \\
C~{\sc ii}& $\ge 3\times10^{14}$       & $\ge 3\times10^{14}$ & \\
N~{\sc i}    & 3.05$\pm0.20\times10^{13}$ & 7.17$\pm0.58\times10^{13}$ &
$-$1.66\\
O~{\sc i}    & $>$6.54$\pm0.12\times10^{15}$ & 1.86$\pm0.12\times10^{15}$ &
$>$$-$0.64\\
Mg~{\sc i}   & 4.85$\pm0.22\times10^{12}$ & 1.68$\pm0.22\times10^{12}$ & \\
Mg~{\sc ii}  & 2.15$\pm0.51\times10^{15}$ & $\le1\times10^{15}$ &
$+$0.05\\
Si~{\sc ii}  & 1.23$\pm0.01\times10^{15}$ & 4.60$\pm0.03\times10^{14}$ &
$-$0.02\\
P~{\sc ii}   & 9.41$\pm0.58\times10^{12}$ & 7.82$\pm0.62\times10^{12}$ &
$-$0.03\\
S~{\sc ii}   & 4.84$\pm0.26\times10^{14}$ & 1.70$\pm0.43\times10^{14}$ &
$-$0.15\\
S~{\sc iii}  & $\le 4\times 10^{14}$      & $\le1\times 10^{15}$ \\
Cr~{\sc ii}  & 3.90$\pm0.17\times10^{12}$ & 3.20$\pm0.25\times10^{12}$ &
$-$0.53\\
Cr~{\sc iii} & $\le 3\times 10^{12}$      & $\le 2\times10^{12}$ \\
Mn~{\sc ii}  & 2.12$\pm0.05\times10^{12}$ & 8.96$\pm0.38\times10^{11}$ &
$-$0.75\\
Fe~{\sc ii}  & 2.40$\pm0.06\times10^{14}$ & 1.20$\pm0.09\times10^{14}$ &
$-$0.65\\
Fe~{\sc iii} & $\le 4\times 10^{14}$      & $\le 5\times 10^{13}$ \\
Ni~{\sc ii}  & 1.85$\pm0.03\times10^{13}$ & 6.15$\pm0.36\times10^{12}$ &
$-$0.56\\
Zn~{\sc ii}  & 7.57$\pm0.48\times10^{11}$ & 4.72$\pm0.60\times10^{11}$ &
$-$0.26\\
\hline
\multicolumn{4}{l}{$^{\rm a}$Metallicty relative to solar,}\\
\multicolumn{4}{l}{[X/H]~=~log~($N$(X)/$N$(H)]$-$log($N$(X)/$N$(H))$_{\odot}$.}
\end{tabular}
\label{tab1}
\end{table}
\subsection{Chemical enrichment}
In what follows we use the standard defition 
[X/H]~=~log~($N$(X)/$N$(H)]$-$log($N$(X)/$N$(H))$_{\odot}$
and the solar metallicities given by Savage \& Sembach (1996).
\par\noindent
Before discussing the chemical composition of the gas, one has to worry 
about potential ionization corections especially for a cloud with 
log~$N$(H~{\sc i})~$\sim$~19.7. From Table~1, it can be seen that 
$N$(Cr~{\sc iii})/$N$(Cr~{\sc ii})~$<$~0.77 and 0.63 and 
$N$(Fe~{\sc iii})/$N$(Fe~{\sc ii})~$<$~1.7 and 0.42 in the two components
respectively. As the two components have otherwise similar characteristics, 
we can conclude that most of the metals are in the singly ionized state.
Howk \& Sembach (1999) have shown that if 
$N$(Fe~{\sc iii})/$N$(Fe~{\sc ii})~$<$~1, the ionization correction for
relative metallicity is negligible for all species of interest here.
Absolute metallicities could be slightly {\sl underestimated} but by no more
than a factor of two. 
\par\noindent
Zinc metallicity is of the order of 0.5 solar. The consistent and mild
depletion (of about a factor of two) of Fe, Cr and Ni compared to Zn
is suggestive of small amounts of dust if any. The four $\alpha$-elements Mg,
Si, P, and S have consistent metallicities slightly larger (by
about a factor of two) than the Zn metallicity. Note however that
sulphur has slightly smaller metallicity compared to the other
$\alpha$-elements. Given the uncertainties involved in
nucleosynthesis models, it is difficult, if not impossible, to draw
any conclusion from one measurement. However it is interesting 
to note that such an abundance pattern is commonly seen in damped
Lyman-$\alpha$ systems (see e.g. Lu et al. 1996) apart from the fact that
the system toward Tol~1037$-$2703 is the damped system of highest metallicity 
at such redshift. In addition, the nitrogen abundance is quite small,
[N/Zn]~=~$-$1.3, [N/S]~=~$-$1.5.
This is broadly consistent with nitrogen being produced as a secondary 
element in massive stars with little contribution from the primary production 
that occurs in intermediate mass stars (3-8 M$\odot$). If true,
this means the star formation in this system was intense and has proceeded 
very quickly through the formation of massive stars. 
However, it is surprizing to find such low [N/Zn]~=~$-$1.3 ratio 
when [Zn/H]~=~$-$0.3 (see e.g. Centurion et al. 1998). 
Note that H~{\sc i} and N~{\sc i} are tied up by charge exchange reaction.
Although one can question the charge exchange reaction rate
between hydrogen and nitrogen especially if the temperature of the gas
is high, it seems difficult to explain the low N~{\sc i}/H~{\sc i} ratio by 
ionization effects only (see also Viegas 1995). Finally, it is most 
interesting to note that the abundance pattern in the  $z_{\rm abs}$~=~2.139 
system is close to what is observed in the Magellanic Bridge (Lehner et al. 
2001) whereas the column densities are similar to those in the 
Weak Low Velocity (WLV) gas toward 23~Orionis (see Welty et al. 1999) except 
for N~{\sc i} which is much stronger toward 23~Ori. 
This could suggest again that this damped Lyman-$\alpha$ system arises
in halo-like gas rather than through a typical disk of galaxy.
\subsubsection {Discussion}
The detection of C~{\sc i} in this system is confirmed by the presence
of lines with $\lambda_{\rm rest}$~=~1650 and 1560 \AA.  
C~{\sc i} has been detected only in few of the intervening damped 
Lyman-$\alpha$ systems [\zabs~=~1.776 toward Q~1331+170 (Songaila et al. 1994);
\zabs~=~1.97 toward Q~0013-00 (Ge et al. 1997); \zabs~=~1.6748 toward
PKS~1756+237 (Roth \& Bauer 1999); \zabs~=~2.337 toward PKS~1232+0815
(Srianand et al. 2000)]. Apart from the system toward PKS 1756+237
(where reliable estimate of the \hi column density is not possible due
to poor S/N in the spectrum) all other systems are confirmed high
column density DLAs. The detection of C~{\sc i} toward Tol~1037$-$2703
is therefore a consequence of the high metallicity of the system.
\par\noindent
Assuming that the carbon abundance is similar to that of silicon and that
C~{\sc ii} is the dominant ionization state we derive a conservative upper 
limit on the C~{\sc ii} column density to be 1.6$\times10^{15}~{\rm cm}^{-2}$ 
in the \zabs~=~2.1394 component. Using the measured column density of
\ciis and assuming collisional excitation and radiative deexcitation
of the fine$-$structure level we determine a range of electron density,
\begin{equation}
n_{\rm e} ~=~(0.48-2.57)~T_4^{0.5}~\exp[91.25/T].
\end{equation}
Assuming photo-ionization equilibrium between C~{\sc i} and {C~\sc ii}
and using the atomic data from Shull \& Van Streenberg (1982), we can write,
\begin{equation}
n_{\rm e}~=~(0.82-4.42)\times 10^9~T_4^{0.64}~\Gamma,  
\end{equation}
where $\Gamma$ is the photo-ionization rate. From the above two equations, 
it can be seen that $\Gamma$~=~$(0.11-4.73)\times 10^{-9}$~s$^{-1}$ 
for a temperature in the range $10^3-10^4$ K.  In the local ISM, 
$\Gamma~\sim~2-3.3\times10^{-10}$ (P\'equignot \& Aldrovandi 1986).
Thus the deduced  ambient UV flux in these components is consistent with
or slightly larger than that in the ISM of our Galaxy.  
\par\noindent
Similarly from Mg~{\sc i} and Mg~{\sc ii} column densities, we obtain
\begin{equation}
n_e~\ge~2.0\times10^9~T_4^{0.89}~\Gamma.
\end{equation}
The allowed value of $\Gamma$ is $\le3.02\times10^{-9}$ when the value in
the Galactic ISM is $\Gamma~=~5.7-8.0\times10^{-11}$. This is consistent
with the previous conclusion.
Note that the hydrogen ionizing background radiation from quasars
and galaxies (Haardt \& Madau 1996) is about two orders of magnitude
smaller than that seen in our galactic ISM and an additional source of
UV flux is needed.  This could be Tol~1037$-$2703 itself as the quasar
is able to provide the additional flux even if the gas is at a
distance $\sim$few Mpc from the QSO. 
\par\noindent
It is known that the fine structure  lines of C~{\sc i} can be excited by the 
cosmic microwave background radiation (Bahcall \& Wolf 1968, Meyer et al. 
1986) and therefore can be used to constrain its temperature. Assuming the 
CMBR pumping to be the only excitation mechanism we obtain an upper limit on 
the temperature of the background to be 16.4 and 13.2 K for \zabs~=~2.1390 
and 2.1394 components respectively (see Srianand et al. 2000 for details on
the derivation). These limits are consistent with the value of 8.6 K 
expected in the framework of standard hot Big-Bang cosmology. 
\section{Summary}
We report the detection of three BAL systems at \zabs~=~1.5181, 1.7018
and 1.7569 toward Tol~1037$-$2703. These systems have apparent ejection 
velocity of, respectively, 36,000, 25,320, and 22,315 \kms~ with respect to 
the accurate emission redshift (\zem~=~2.201) we measure using the Mg~{\sc ii}
emission line. The identification of the former BAL is based on Mg~{\sc i},
Mg~{\sc ii}, Fe~{\sc ii} and Al~{\sc ii} absorptions and the system
belongs to the class of low ionization Mg~{\sc ii} BALs. The latter two 
systems are typical high-ionization BALs with absorptions from \nv and \civ 
and no low-ionization lines. Thus Tol~1037$-$2703 is unique in the sense
that this is the only QSO known that shows well detached BALs with widely
differing ionization conditions. Broadness of the profiles and
inferred covering factors suggest that the projected size of the corresponding
absorbing clouds is less than that of the UV continuum emitting region. 
\par\noindent
We show, using \civ and \aliii doublets, that the gas responsible for
the \zabs~=~2.082 system does not cover the background source of continuum
completely. Spectra of Tol~1037$-$2703 obtained at two different
epochs, separated by $\sim$2 yrs in the rest frame of the QSO, suggests
variability of the \civ absorption. The relative strength of
the fine-structure lines due to \siii and \cii is a consequence of excitation
by a strong IR source, probably associated with the quasar. 
Similar properties have been derived in the case of the $z_{\rm abs}$~=~3.8931
system toward APM~08279$+$5255 (Srianand \& Petitjean 2000). 
We note that the covering factor for \civ, $f_{\rm c}$~$\sim$~0.86, is 
constant over the velocity range span by the absorption line. The covering 
factor estimated for Al~{\sc iii}, $f_{\rm c}$~$\sim$~0.3, is much less, 
confirming the findings by Petitjean \& Srianand (1999) that the covering 
factor is increasing with increasing ionization state. 
\par\noindent
We report the detection of weak low-ionization lines such as C~{\sc i}, 
Zn~{\sc ii}, Cr~{\sc ii}, Ni~{\sc ii}, P~{\sc ii} and S~{\sc ii}
in the moderately damped (log~$N$(H~{\sc i})~=~19.7) \zabs~2.139 system.  
These lines are detected in two
components at \zabs~2.1390 and 2.1394. The absolute metallicity is
close to solar with [Zn/H]~=~$-$0.26. The $\alpha$-chain elements have
metallicities consistently solar (respectively +0.05, $-$0.02, $-$0.03
and $-$0.15 for [Mg/H], [Si/H], [P/H] and [S/H]) and iron peak
elements are depleted by a factor of about two ([Fe/Zn], [Cr/Zn],
[Mn/Zn] and [Ni/Zn] are equal to $-$0.39, $-$0.27, $-$0.49,
$-$0.30). It is noticeable that the pattern is similar to what is seen
in standard damped Lyman-$\alpha$ systems (Lu et al.  1996). However,
in the case of Tol~1037$-$2703, not only is the metallicity high but
also the nitrogen abundance is small, [N/Zn]~=~$-$1.40 which is
difficult to understand in standard nucleosynthesis models. 
Such metallicity pattern is seen in the Magellanic Bridge however
(Lehner et al. 2001). In any case this clearly indicates that the period of 
intense star formation activity in this system has last for less than 
0.5~$Gyr$ and that metals have primarily been produced by massive stars. 
Using the C~{\sc ii$^*$} column density, the Mg~{\sc i}/Mg~{\sc ii} and 
C~{\sc i}/C~{\sc ii} ratios, we show that the required photo-ionization rates 
are higher than what is expected if the gas is ionized by the UV background 
and some extra source of ionization is needed. Assuming that the QSO is
the ionizing source implies that the absorbing gas is situated
$\sim$few Mpc away from it. The high metal enrichment, within few Mpc from 
a QSO that, in addition, shows a large variety of outflowing gas might 
suggest a possible association of this gas with the QSO. It is possible that 
this absorbing gas has in common with the quasar a period of intense star 
formation and/or has been ejected from the QSO few 10$^8$ yrs back.
\par\noindent
Using the limits on the C~{\sc i$^*$} column density and assuming excitation 
is due to CMBR photons we obtain an upper limit on the temperature of the 
cosmic microwave background of 16.2 and 13.2 K respectively for these 
components. These limits are consistent with the expected temperature of 
8.6 K in the frame work of standard hot Big$-$bang cosmology.
%
%%%%%%%%%%%%%%%%%%%%%%%%%%%%%
\acknowledgement{We thank C\'edric Ledoux for a careful reduction 
of the spectrum and Chris Impey for sharing with us a 1994 MMT spectrum. 
We gratefully acknowledge support from the Indo-French Centre for 
the Promotion of Advanced Research (Centre Franco-Indien pour la Promotion
de la Recherche Avanc\'ee) under contract No. 1710-1. RS thanks IAP
for hospitality during the time part of this work was completed.
}
 
%%%%%%%%%%%%%%%%%%%%%%%%%%%%%

\end{document}